\documentclass{aa}
\usepackage[varg]{txfonts}
\usepackage{bm}
\usepackage{color}
\usepackage{natbib}
\usepackage{amssymb}
\usepackage{ulem}

\usepackage{etoolbox}
\usepackage{orcidlink}

\bibpunct{(}{)}{;}{a}{}{,} 

\newcommand{\rmd }{{\rm d}}  
\newcommand{\rmi }{{\rm i}}  
\newcommand{\adi}{ARDI}

\begin{document}

\title{Resonant Drag Instabilities for Polydisperse Dust}

\subtitle{I. The Acoustic Resonant Drag Instability}

\author{
Sijme-Jan Paardekooper\orcidlink{0000-0002-8378-7608}\inst{1}
\and Hossam Aly\orcidlink{0000-0002-1342-1694}\inst{1}
}

\institute{Faculty of Aerospace Engineering, Delft University of Technology, Delft,The Netherlands}

\date{Received *** / Accepted ***}

\abstract {Dust grains embedded in gas flow give rise to a class of hydrodynamic instabilities that can occur whenever there exists a relative velocity between gas and dust. These instabilities have predominantly been studied for single grain sizes, for which a strong interaction can be found between drifting dust and a travelling gas wave, leading to fast-growing perturbations (growth rates $\propto \sqrt{\mu}$) even at small dust-to-gas ratios $\mu$. They are called resonant drag instabilities. We focus on the acoustic resonant drag instability, which is potentially important in AGB star outflows, around supernova remnants and star clusters in starburst galaxies.} 
{We study the acoustic resonant drag instability, taking into account a continuous spectrum of grain sizes, to determine whether it survives in the polydisperse regime and how the resulting growth rates compare to the monodisperse case.} 
{We solve the linear equations for a polydisperse fluid for the acoustic drag instability, focusing on small dust-to-gas ratios. } 
{Size distributions of realistic width turn the fast-growing perturbations $\propto \sqrt{\mu}$ of the monodisperse limit into slower growing perturbations $\propto \mu$ due to the fact that the backreaction on the gas involves an integration over the resonance. Furthermore, the large wave numbers that grow fastest in the monodisperse regime are stabilized by a size distribution, severely limiting the growth rates in the polydisperse regime.} 
{The acoustic resonant drag instability turns from a singularly perturbed problem in $\mu$ in the monodisperse limit into a regular perturbation for a sufficiently wide size distribution. It can still grow exponentially in the polydisperse regime, but at a slower pace compared to the single size case.}


\maketitle 

\section{Introduction}

Visible matter in the universe consists mostly of gas, $0.99$ being the canonical fraction \citep[see e.g.][]{2007ApJ...663..866D}. The remaining $1\%$ is in solid form, and consists of particles of various sizes, from sub-micron size to planetary size, that are often mixed with the gas. Examples of dusty gas flows include AGB star winds \citep{2018A&ARv..26....1H}, the interstellar medium \citep{2003ARA&A..41..241D}, and protoplanetary discs \citep{2023ASPC..534..423B}. The dust component can play an important role in the chemistry and opacity of the mixture, for example, but also in its dynamics. In particular, adding dust to a gas in an otherwise stable configuration can make the system go hydrodynamically unstable. 

One particular class of instabilities in gas-dust mixtures are the so-called Resonant Drag Instabilities \citep[RDIs, see][]{2018MNRAS.477.5011S, 2018ApJ...856L..15S}. These instabilities rely on the fact that gas and dust are usually drifting relative to each other. If a wave in the gas can propagate at the same velocity as the drift velocity of the dust, the result is often a very strong instability that grows fast even if the total amount of dust is very small \citep{2018ApJ...856L..15S}. Of particular interest is the Streaming Instability \citep[SI,][]{2005ApJ...620..459Y}, which can play an important role in planet formation, and was shown to be an inertial wave RDI \citep{2018MNRAS.477.5011S, 2019MNRAS.489.3850Z, 2024MNRAS.534.3944M}. If the system of interest allows for different gas waves, different types of RDI can manifest, such as the acoustic RDI \citep{2018MNRAS.480.2813H} or RDIs in magnetized gas \citep{2020MNRAS.496.2123H}.  

Dust drift speeds usually depend on particle size. When accelerated for example due to gravity, dust particles will eventually reach an equilibrium terminal velocity where the acceleration is balanced by (size-dependent) gas drag. The smallest particles, tightly coupled to the gas, will drift slowest, while larger particles will drift faster. Since RDIs rely on the drift speed being equal to the gas wave speed, it is natural to ask what happens to an RDI when there is a distribution of dust sizes, and therefore a distribution of drift velocities.

In this paper we focus on the acoustic resonant drag instability. This instability occurs in systems where drift is fast enough to couple to sound waves, and where magnetic forces are subdominant \citep[see figure 2 of ][]{2022MNRAS.517.1491H}. A recent physical picture of the acoustic RDI was presented in \cite{2024MNRAS.529..688M}. The instability occurs for example in dust-driven outflows, where radiation pressure drives dust particles away from AGB stars \citep[e.g.][]{2018A&ARv..26....1H}, around supernova remnants \citep{2018SSRv..214...53M}, or around star clusters in starburst galaxies \citep{2023MNRAS.521.5160M}. The nonlinear outcome of the acoustic RDI can be virulent turbulence with strong dust clumping \citep{2019MNRAS.489..325M}. This will have strong implications for the chemistry and opacity in the system \citep[see also][]{2019MNRAS.485.3991S,2020MNRAS.496.2123H}. 

The purpose of this paper is twofold: first of all, we want to study the linear phase of the acoustic RDI with a dust size distribution for its own sake. In this sense, we are continuing the work of \cite{2022MNRAS.510..110S}, who mostly focused on the case where all sizes have the same drift velocity (but see their appendix a3). We will deal exclusively with the case where the drift speed is different for different grain sizes. The second purpose of this paper is to provide a framework for interpreting RDIs in the polydisperse regime. Since the acoustic RDI (hereafter \adi{}) is probably the simplest RDI (in terms of physics required as well as the number of spatial dimensions), it lends itself well for this. In a companion paper, we use the framework to deal with more complex cases of the streaming instability \citep{2005ApJ...620..459Y} and settling instability \citep{2018ApJ...856L..15S, 2018MNRAS.477.5011S}, which are likely important in protoplanetary discs during planet formation. 

The plan of this paper is as follows. We introduce the basic equations in section \ref{sec:eq_general}, where we also define our equilibrium state and derive equations for linear perturbations. Results are presented in section \ref{sec:results}, and we conclude in section \ref{sec:conclusion}.

\section{Governing equations}
\label{sec:eq_general}

The equations governing a mixture of gas and dust, where the dust component consist of a continuum of sizes, were presented in \cite{2021MNRAS.502.1579P}:
\begin{align}
\partial_t\sigma 
+ \nabla\cdot(\sigma {\bf{u}} )=& 0,\label{eq:dustcont}\\
\partial_t{\bf{u}}
+ ({\bf{u}}\cdot\nabla){\bf{u}}
=&
\bm{\alpha}_{\rm d}  - \frac{{\bf{u}} - {\bf{v}_{\rm g}}}{{\tau_{\rm s}}}.\label{eq:dustmom}
\end{align}
Here, $\sigma$ the size density \citep{2020MNRAS.499.4223P}, ${\bf u}$ the size-dependent dust velocity, and ${\bf{v}_{\rm g}}$ the gas velocity. We specialize drag law of the form $({\bf{u}}-{\bf{v}_{\rm g}})/{\tau_{\rm s}}$ to couple gas and dust, with stopping time ${\tau_{\rm s}}$ \citep[see e.g.][]{2003A&A...399..297W}. While the stopping time can in principle depend on gas density and $|{\bf{u}}-{\bf{v}_{\rm g}}|$ \citep{2018ApJ...856L..15S}, we will consider only the simple case where the stopping time does not depend on any dynamical quantity. Any additional accelerations acting on the dust are contained in $\bm{\alpha}_{\rm d}$, which we assume to depend on ${\bf{u}}$ only. Note that under this assumption, the dust momentum equation is completely decoupled from the evolution of the size density. 

The gas component, similarly, has its continuity and momentum equation, where the backreaction of drag on the gas enters the latter:
\begin{align}
\partial_t \rho_{\rm g} + \nabla\cdot (\rho_{\rm g} {\bf v}_{\rm g})=&0,\label{eq:gascont}\\
\partial_t{\bf{v}_{\rm g}} + ({\bf{v}_{\rm g}}\cdot\nabla){\bf{v}_{\rm g}} =& -\frac{\nabla p}{{\rho_{\rm g}}} +\bm{\alpha}_{\rm g} + \frac{1}{{\rho_{\rm g}}}\int \sigma \frac{{\bf{u}} - {\bf{v}_{\rm g}}}{{\tau_{\rm s}}}\rmd{\tau_{\rm s}}\, ,
\label{eq:gasmom}
\end{align}
where ${\rho_{\rm g}}$ is the gas density, $p$ the pressure, and all additional accelerations on the gas are contained in $\bm{\alpha}_{\rm g}$. At this point, we can remain agnostic about any other equations governing the gas; there might be an energy equation, equations coupling the magnetic field, etc. The only assumption here is that the only way gas and dust interact is through the drag force. Nevertheless, in this paper, we will be concerned with only very simple gas physics: no magnetic fields, and a barotropic equation of state, so that (\ref{eq:gascont}) and (\ref{eq:gasmom}) completely describe the evolution of the gas component. 

It is worth noting that, as long as we do not consider any interaction between the dust particles themselves, we have complete freedom in letting $\sigma$ depend on multiple dust properties. Apart from size, one could consider shape, porosity, chemical composition, etc. Given dust parameters $p_1$, $p_2$, $\dots$, $p_n$, we write
\begin{align}
\sigma =& \sigma(p_1, p_2, \dots, p_n),\\
{\tau_{\rm s}} =& {\tau_{\rm s}}(p_1, p_2, \dots, p_n),
\end{align}
and demand that
\begin{align}
{\rho_{\rm d}}=\int \sigma \rmd p_1\rmd p_2 \dots \rmd p_n,
\end{align}
where ${\rho_{\rm d}}$ is the mass volume density of dust. The drag integral appearing in (\ref{eq:gasmom}) will also be over all dust parameters. This will be useful if for example $\bm{\alpha}_{\rm d}$ depends on dust parameters, or if one considers a more complex drag law. In this paper, we do not consider these cases, but since the only relevant dust parameter is the stopping time, we can simplify notation by choosing $p_1={\tau_{\rm s}}$, so that the drag integral is performed over stopping time rather than size\footnote{Technically, $\sigma$ should then be called 'stopping time density', but note that for spherical particles the stopping time is proportional to the particle size.}. 

The size distribution is often taken to be a power law, at least initially. Most often, the MRN distribution is considered \citep{1977ApJ...217..425M}, which has the number density of dust grains $\propto a^{-7/2}$, where $a$ is the dust size, and therefore $\sigma\propto {\tau_{\rm s}}^{-1/2}$. Different power laws were explored in e.g. \cite{2019ApJ...878L..30K, 2021MNRAS.501..467Z}, while in \cite{2021MNRAS.502.1469M} power laws were augmented with a bump at larger sizes to account for grain growth and in particular cratering \citep{2011A&A...525A..11B}. In this paper, to make the various integrals more tractable, we will concern ourselves with very simple size distributions only. In particular, we will study a constant size distribution between a minimum and a maximum stopping time:
\begin{align}
    \sigma({\tau_{\rm s}}) = \left\{\begin{array}{ll} \frac{{\rho_{\rm d}}}{{\tau_{\rm s,max}}-{\tau_{\rm s,min}}} & {\tau_{\rm s,min}} < {\tau_{\rm s}} < {\tau_{\rm s,max}}\\ 0 & {\rm otherwise} \end{array}\right.
    \label{eq:dimsizedist}
\end{align}
It is straightforwardly verified that the integral over stopping time yields the total dust density, ${\rho_{\rm d}}$. 

\subsection{Equilibrium state}

The \adi{} can be set up in 1D Cartesian geometry. We fix the gas equation of state to be isothermal with constant sound speed $c_{\rm g}$, and consider a drag law where the stopping time $\tau_{\rm s}$ is independent of gas density. The dust size distribution is parametrized using the size density $\sigma(\tau_{\rm s})$, defined so that the dust density is
\begin{align}
  \rho_{\rm d}= \int \sigma(\tau_{\rm s}) {\rm d} \tau_{\rm s}.
\end{align}
The equations governing the dynamics of the coupled dust-gas system are given by conservation of mass and momentum of gas and dust:
\begin{align}
  \partial_t\rho_{\rm g} + \partial_x(\rho_{\rm g} v_{\rm g})=&0,\\
  \partial_t v_{\rm g} + v_{\rm g} \partial_xv_{\rm g} =& -\mu \alpha - \frac{c_{\rm g}^2\partial_x\rho_{\rm g}}{\rho_{\rm g}} + \frac{1}{\rho_{\rm g}}\int \frac{\sigma}{\tau_{\rm s}}(u-v_{\rm g}){\rm d} \tau_{\rm s},\\
   \partial_t\sigma + \partial_x(\sigma u)=&0,\\
  \partial_t u+ u\partial_xu =& \alpha-\frac{u-v_{\rm g}}{\tau_{\rm s}},
\end{align}
where the dust velocity $u=u(\tau_{\rm s})$. A constant acceleration $\alpha$ was chosen to generate drift, similar to \cite{2018MNRAS.480.2813H} and \cite{2024MNRAS.529..688M}. In reality, this may for example come from radiation pressure, see e.g. \cite{2022MNRAS.515.4797S}. An equilibrium state can be constructed where all quantities are constant in space and time, with drift velocity $u^{(0)}-v_{\rm g}^{(0)}=\alpha\tau_{\rm s}$. For simplicity, we choose $v_{\rm g}^{(0)}=0$, or, equivalently, we work in a frame that moves with the background gas velocity. We denote the equilibrium gas density $\rho_{\rm g}^{(0)}$, equilibrium dust size density $\sigma^{(0)}({\tau_{\rm s}})$, and equilibrium dust to gas ratio $\mu = \int\sigma^{(0)}\rmd{\tau_{\rm s}}/\rho_{\rm g}^{(0)}$.

\subsection{Linear perturbations}

Consider small perturbations around the equilibrium state, denoting perturbed quantities with superscript '1':
\begin{align}
  \partial_t\rho_{\rm g}^{(1)} =& -\rho_{\rm g}^{(0)}\partial_xv_{\rm g}^{(1)},\\
  \partial_t v_{\rm g}^{(1)} =& -\frac{c_{\rm g}^2\partial_x\rho_{\rm g}^{(1)}}{\rho_{\rm g}^{(0)}} + 
  \frac{1}{\rho_{\rm g}^{(0)}}\int \frac{\sigma^{(0)}}{\tau_{\rm s}}(u^{(1)}-v_{\rm g}^{(1)}){\rm d} \tau_{\rm s}+ \nonumber\\
  &\int \left(\frac{\sigma}{\rho_{\rm g}}\right)^{(1)}\frac{u^{(0)}-v_{\rm g}^{(0)}}{\tau_{\rm s}}{\rm d} \tau_{\rm s},\\
   \partial_t\sigma^{(1)} =&-\sigma^{(0)}\partial_xu^{(1)} - u^{(0)}\partial_x\sigma^{(1)},\\
  \partial_t u^{(1)} =&-u^{(0)}\partial_xu^{(1)} -\frac{u^{(1)}-v_{\rm g}^{(1)}}{\tau_{\rm s}}.
\end{align}
Take perturbations $\rho_{\rm g}^{(1)} = \hat\rho_{\rm g} \exp({\rm i}kx - {\rm i}\omega t)$, and similar for other perturbed quantities :
\begin{align}
  \omega {\hat{\rho}_{\rm g}} =& k\rho_{\rm g}^{(0)}\hat v_{\rm g},\label{eq:pertgasdens}\\
  \omega \hat{v}_{\rm g} =& \frac{k c_{\rm g}^2\hat \rho_{\rm g}}{\rho_{\rm g}^{(0)}} + 
  \frac{\rmi}{\rho_{\rm g}^{(0)}}\int \frac{\sigma^{(0)}}{\tau_{\rm s}}(\hat u-\hat v_{\rm g}){\rm d} \tau_{\rm s}+ \nonumber\\
  &\rmi \int \hat{\left(\frac{\sigma}{\rho_{\rm g}}\right)}\frac{u^{(0)}-v_{\rm g}^{(0)}}{\tau_{\rm s}}{\rm d} \tau_{\rm s},\\
   \omega \hat\sigma =& k\sigma^{(0)}\hat u + k u^{(0)}\hat \sigma,\label{eq:dustpert}\\
  \omega \hat u =& k u^{(0)}\hat u -\rmi \frac{\hat u-\hat v_{\rm g}}{\tau_{\rm s}},
  \label{eq:pertdustvel}
\end{align}
where dust quantities $\hat\sigma$ and $\hat u$ depend on the stopping time. Note that the perturbation in dust to gas ratio
\begin{align}
  \hat{\left(\frac{\sigma}{\rho_{\rm g}}\right)}=\frac{\hat\sigma}{\rho_{\rm g}^{(0)}} - \frac{\sigma^{(0)}}{\rho_{\rm g}^{(0)}}\frac{\hat \rho_{\rm g}}{\rho_{\rm g}^{(0)}},
\end{align}
so that
\begin{align}
  {\rm i}\int \hat{\left(\frac{\sigma}{\rho_{\rm g}}\right)}\frac{u^{(0)}-v_{\rm g}^{(0)}}{\tau_{\rm s}}{\rm d} \tau_{\rm s} =  {\rm i}\alpha\int \frac{\hat\sigma}{\rho_{\rm g}^{(0)}} {\rm d} \tau_{\rm s}-  {\rm i}\alpha\mu\frac{\hat \rho_{\rm g}}{\rho_{\rm g}^{(0)}}.
\end{align}
Equations (\ref{eq:pertgasdens})-(\ref{eq:pertdustvel}) form a continuous eigenvalue problem in ${\tau_{\rm s}}$, with eigenvalue $\omega$, with a linear operator involving integral terms. These integrals can cause significant difficulties when determining eigenvalues \citep{2021MNRAS.502.1579P}. These difficulties are related to the presence of the resonance that in the monodisperse regime gives the RDIs their name. In the case of the \adi{}, we see from (\ref{eq:dustpert}) that if $\omega=ku^{(0)}$ the dust density perturbation diverges. If in the limit of $\mu\rightarrow 0$ we are dealing with a gas sound wave, i.e. $\omega=kc_{\rm g}$, the resonant condition requires dust to drift at the sound speed. This is where fast growth in the monodisperse regime is expected \citep[e.g.][]{2018ApJ...856L..15S, 2018MNRAS.477.5011S}. In the polydisperse case, we often have to integrate over this resonance, the presence of which makes the integral nearly singular \citep[for examples see][see also the discussion below equation (\ref{eq:adi_wide_integral})]{{2021MNRAS.502.1579P}}. This is where standard integration techniques fail \citep{2021MNRAS.502.1579P}.

\subsection{Dispersion relation}

We can combine (\ref{eq:pertgasdens})-(\ref{eq:pertdustvel}) into a single equation for the temporal frequency $\omega$:
\begin{align}
  \omega^2 - c_{\rm g}^2 k^2 + \int \frac{\sigma^{(0)}}{\rho_{\rm g}^{(0)}}\frac{\omega\left(\omega- u^{(0)}  k\right)}{1-{\rm i}\tau_{\rm s}\left(\omega- u^{(0)}  k\right)}{\rm d} \tau_{\rm s}- \nonumber\\
  {\rm i}\alpha\int \frac{\sigma^{(0)}}{\rho_{\rm g}^{(0)}}\frac{k\omega}{\omega-ku^{(0)}}\frac{1}{1-{\rm i}\tau_{\rm s}\left(\omega- u^{(0)}  k\right)} {\rm d} \tau_{\rm s}+  {\rm i}\alpha k\mu=0,
\end{align}
The monodisperse limit is obtained by taking $\sigma^{(0)}$ to be a Dirac delta function at stopping time $\tau_{\rm s}$:
\begin{align}
  \left(\omega^2-k^2 c_{\rm g}^2\right) \left(\omega- u^{(0)}  k\right)\left(\frac{{\rm i}}{\tau_{\rm s}}+\omega- u^{(0)}  k\right)  +\nonumber\\
              \frac{{\rm i}\mu}{\tau_{\rm s}}\left[
              (\omega+\alpha k\tau_{\rm s})\left(\omega- u^{(0)}  k\right)^2
              - {\rm i} \alpha k^2u^{(0)}
              \right]=0,
            \label{eq:adi_mono_disp}
\end{align}
which agrees with \cite{2024MNRAS.529..688M}. A non-dimensional version, with time scale $c_{\rm g}/\alpha$ and length scale $c_{\rm g}^2/\alpha$, reads:
\begin{align}
  \tilde\omega^2 =&
              K^2 + \mu
              \int \Sigma(s)\frac{{\rm i}K\tilde\omega-\tilde\omega(\tilde\omega-Ks)^2}{(\tilde\omega-Ks)(1-{\rm i}s\left(\tilde\omega- Ks\right))}{\rm d} s - {\rm i}K\mu,
\label{eq:adi_disp}
\end{align}
with $\tilde\omega=c_{\rm g}\omega/\alpha$, $K=kc_{\rm g}^2/\alpha$, $s = \tau_{\rm s} \alpha/c_{\rm g}$ and a nondimensional size distribution $\Sigma = c_{\rm g}\sigma^{(0)}/(\alpha\rho_{\rm d}^{(0)})$, so that $\int\Sigma(s){\rm d}s=1$. It is worth noting the various contributions on the right hand side: pressure, dust density (first term in the numerator under the integral), relative velocity (second term in the numerator under the integral), and gas density (last term, outside the integral). The stability of the system is then determined by $\mu$, $K$, and the size distribution. The non-dimensional version of the size density (\ref{eq:dimsizedist}) that we will use for the \adi\, reads:
\begin{align}
  \Sigma(s) = \left\{\begin{array}{ll}
                       \frac{1}{s_{\rm max}-s_{\rm min}} & s_{\rm min} < s < s_{\rm max} \\
                       0 & {\rm otherwise.}
                     \end{array}\right.
                     \label{eq:sizedistsigma}
\end{align}

Numerical results were obtained by solving (\ref{eq:adi_disp}) with the techniques detailed in \cite{2021MNRAS.502.1579P,2021zndo...4305344M}.

\section{Linear \adi{} modes}
\label{sec:results}

In the analysis of resonant drag instabilities, the dust to gas ratio $\mu$ is assumed to be small, and a series solution is developed. From (\ref{eq:adi_mono_disp}), we see that in the monodisperse case, there are two possible (non-damped) solutions in the limit $\mu \rightarrow 0$: a gas sound wave with $\omega^2=k^2c_{\rm g}^2$, and a dust advection mode with $\omega=u^{(0)}k$. Since the gas does not feel the dust for $\mu=0$, the sound wave propagates independent of any dust perturbation, while for the dust advection mode, a dust density perturbation is advected with the unperturbed dust velocity $u^{(0)}$, without any dust velocity perturbation. 

\subsection{Non-resonant growth}
\label{sec:nonresonant}

We first focus on the non-resonant case, when dust drift is different from the gas sound speed \citep[see also][]{2018MNRAS.480.2813H}. If $\tilde\omega=\pm K \neq Ks$, we can have a non-resonant growing sound wave in the monodisperse regime. Trying a solution $\tilde\omega = K + \mu \tilde\omega_1$ in (\ref{eq:adi_mono_disp}) yields
\begin{align}
  \tilde\omega_1  =
              \frac{1}{2}\frac{
              K(1+s)\left(s-1\right)^2
              - {\rm i} s
              }{\left(s-1\right)\left(1+\rmi Ks \left(s-1\right)\right)},
            \label{eq:ardi_nr_sound}
\end{align}
from which we can deduce that growing modes exist when $s < 1$ and
\begin{align}
    K^2 < \frac{1}{(s+1)\left|s-1\right|^3}.
    \label{eq:adi_nonres_cutoff}
\end{align}
Note that, for our ordering in $\mu\ll 1$ to be valid, we can not have $K$ smaller than $\mu$. 
Interestingly, these growing modes can survive in the polydisperse case. In the limit of $K\rightarrow 0$, with a size distribution (\ref{eq:sizedistsigma}) such that $s_{\rm max}<1$, we find that 
\begin{align}
    \Im(\tilde \omega_1)=-\frac{1}{2}\left(\log\left(\frac{1-s_{\rm max}}{1-s_{\rm min}}\right)+1\right).
\end{align}
Physically, we get growth if the destabilising effect of the dust density perturbation, represented by the logarithm, overcomes the stabilising effect of the gas density perturbation (the constant term, originating from the last term in (\ref{eq:adi_disp})). This can happen if $s_{\rm max}$ is close enough to the resonant size $s=1$. For finite $K$, we get an additional stabilising effect from the dust velocity perturbation, which means we may need to approach the resonant size even closer. From (\ref{eq:ardi_nr_sound}), it is clear that the approximation breaks down at the resonant size where $s=1$, but as long as we stay away from this resonance, we can have growing modes $\tilde \omega = K + \mu \tilde \omega_1$. The condition for being far enough away from the resonance is simply $|\tilde\omega_1| \ll K/\mu$. 

If we choose a backward propagating sound wave we can avoid the resonance altogether \citep[see also][]{2022MNRAS.510..110S}. Trying a solution $\tilde\omega=-K + \mu\tilde\omega_1$ in (\ref{eq:adi_disp}), we find that 
\begin{align}
    \tilde\omega_1=\frac{\rmi}{2} + \frac{1}{2}\int \Sigma(s)\frac{K(1+s)^2 - \rmi}{(1+s)(1+\rmi Ks(1+s))}\rmd s.
    \label{eq:adi_backward}
\end{align}
The first term is due to the gas density perturbation, while the first term in the numerator under the integral sign is due to the relative velocity perturbation, and the second term is due to the dust density perturbation. The integrand is well-behaved for $s>0$, and while the closed form solution does not provide much insight, a numerical solution is straightforward to calculate. In the limits of small and large $K$, $\omega_1$ becomes independent of $K$:
\begin{align}
    \tilde\omega_1=\frac{\rmi }{2}
    \left\{\begin{array}{cc}1-\frac{1}{s_{\rm max}-s_{\rm min}}\log\left(\frac{1+s_{\rm max}}{1+s_{\rm min}}\right), & K\ll 1\\
    1-\frac{1}{s_{\rm max}-s_{\rm min}}\log(s_{\rm max}/s_{\rm min}) & K \gg 1. 
    \end{array}\right.   
    \label{eq:adi_backward_approx}
\end{align}
For $K \ll 1$, we always get growth for our constant size distribution, while for $K \gg 1$ we usually have $\Im(\tilde\omega_1) < 0$, as long as $s_{\rm min}$ is sufficiently smaller than unity. The transition from growth to damping at $K\sim 1$ was also found by \cite{2022MNRAS.510..110S}. As a reminder, note that we need $K>\mu$ for our ordering scheme to remain valid. 

If instead of a sound wave we choose a non-resonant dust advection mode with $\tilde\omega=Ks \neq K$, an expansion $\tilde\omega = Ks + \mu \tilde\omega_1$ yields in the monodisperse case
\begin{align}
  \tilde \omega_1  = {\rm i} \frac{s}{s^2-1},
\end{align}
with growth for $s > 1$. In this case, however, the generalisation to the polydisperse case is less straightforward. Consider an expansion $\tilde\omega = Ks_* + \mu\tilde\omega_1$ for some dimensionless, non-resonant stopping time $s_*$. Equation (\ref{eq:adi_disp}) gives, up to first order in $\mu$:
\begin{align}
2\mu Ks_*\tilde\omega_1=
              K^2(1-s_*^2) + \nonumber\\
              \mu
              \int \Sigma(s)\frac{{\rm i}K\tilde\omega-\tilde\omega(\tilde\omega-Ks)^2}{(Ks_*+\mu\tilde\omega_1-Ks)(1-{\rm i}s\left(\tilde\omega- Ks\right))}{\rm d} s - {\rm i}K\mu,
\end{align}
where for clarity of notation we have only expanded the relevant terms of $\tilde\omega$. At zeroth order in $\mu$, in order to have a non-resonant mode with $s_* \neq 1$, the first term on the right hand side needs to be balanced by a term in the integral. This happens if $K|s-s_*|\ll \mu \tilde \omega_1$ for all $s$. In this case, we simply find at zeroth order in $\mu$ that
\begin{align}
K^2(1-s_*^2) +
              \int \Sigma(s)\frac{{\rm i}K^2s_*}{\tilde\omega_1}{\rm d} s  = 0,
\end{align}
which immediately leads to the equivalent of the monodisperse result 
\begin{align}
    \tilde\omega_1=\frac{\rmi s_*}{s_*^2-1},
\end{align}
with growth for $s_*>1$, which corresponds to supersonic drift. The condition that $K|s-s_*|\ll \mu \tilde \omega_1$ is a condition on the width of the size distribution: away from the resonance, $\tilde\omega_1=O(1)$, so that we need $|s-s_*|\ll \mu/K$. Only for such a very narrow size distribution does the dust advection mode exist (remember $\mu \ll 1$): if the size distribution is too wide, there can be no coherent gas reaction to sustain the mode. This was also noted in \cite{2020MNRAS.499.4223P} for the streaming instability, which, away from the resonance, also relies on a dust advection mode to go unstable \citep{2005ApJ...620..459Y}.

\subsection{Resonant modes}

We now turn to the resonant case. As was shown in \cite{2018ApJ...856L..15S}, if a gas wave propagates with the same velocity as the dust is drifting, a very strong reaction results even for $\mu \ll 1$. Mathematically, this is manifested as a growth rate proportional to $\sqrt{\mu}$ instead of $\mu$, which for $\mu \ll 1$ usually leads to stronger growth. In other words, at resonance, the perturbation of the system in the small parameter $\mu$ is singular, and in the limit of $\mu \rightarrow 0$ multiple distinct solution collapse onto each other. The associated defective nature of the matrix eigenvalue problem lies at the heart of the resonant drag instability formalism \citep{2018ApJ...856L..15S}.

Following this reasoning, we  try an expansion $\tilde\omega = K(1 + \epsilon\tilde\omega_1 + \epsilon^2\tilde\omega_2)$, with $\mu = \epsilon^2\bar\mu$ with $\bar\mu=O(1)$:
\begin{align}
 2\epsilon\tilde\omega_1 + \epsilon^2(\tilde\omega_1^2+2\tilde\omega_2) -\nonumber\\
              \frac{\bar\mu\epsilon^2}{K}\int \frac{\Sigma(s)\left({\rm i}-K(1 + \epsilon\tilde\omega_1 + \epsilon^2\tilde\omega_2-s)^2\right)}{(1 + \epsilon\tilde\omega_1 + \epsilon^2\tilde\omega_2-s)(1-{\rm i}sK\left(1 + \epsilon\tilde\omega_1 + \epsilon^2\tilde\omega_2- s\right))}{\rm d} s \nonumber\\ + \frac{{\rm i}\bar \mu\epsilon^2}{K}=0,
\end{align}
where we formally assume $K=O(1)$. Note that in the absence of dust ($\mu=0$), we have a sound wave with $\tilde \omega=K$, or, equivalently, $\omega = k c_{\rm g}$. The resonant condition for an RDI is $\tilde\omega = Ks$, which means at the resonance we need $s=1$, which corresponds to exactly sonic drift. We first make contact with the monodisperse results by considering a narrow size distribution.

\subsubsection{Narrow size distribution}
Consider a narrow size distribution (to be defined below) around the resonance at $s=1$. Under the integral, we only need to keep terms up to order $\epsilon$. Let
\begin{align}
  I\equiv & \epsilon^2\int \frac{\Sigma(s)\left({\rm i}-K(1 + \epsilon\tilde\omega_1-s)^2\right)}{(1 + \epsilon\tilde\omega_1-s)(1-{\rm i}sK\left(1 + \epsilon\tilde\omega_1 - s\right))}{\rm d} s\nonumber\\ 
  =&\epsilon I^{(1)} + \epsilon^2I^{(2)},
\end{align}
allowing for the singularity at $s=1$ to lead to a term $I^{(1)}$. It is convenient to split the integral:
\begin{align}
  I\equiv {\rm i}\epsilon^2\int \frac{\Sigma(s)}{(1 + \epsilon\tilde\omega_1 -s)(1-{\rm i}sK\left(1 + \epsilon\tilde\omega_1 - s\right))}{\rm d} s -\nonumber\\
\epsilon^2K\int \frac{\Sigma(s)(1-s+{\rm i}sK(1-s)^2)}{1+s^2K^2\left(1 - s\right)^2}{\rm d} s.
\end{align}
The first term is due to the dust density perturbation, the second to the perturbation in relative velocity. The last term is $O(\epsilon^2)$ for all values of $K$. The first integral can be analysed by noting the resonance at $s=1$, and developing a series around $s=1$:
\begin{align}
  {\rm i}\epsilon^2\int \frac{\Sigma(s)}{(1 + \epsilon\tilde\omega_1 -s)(1-{\rm i}sK\left(1 + \epsilon\tilde\omega_1 - s\right))}{\rm d} s = \nonumber\\
  {\rm i}\epsilon^2\int \frac{\sum_{n=0}^\infty A_n (1-s)^n}{1 + \epsilon\tilde\omega_1 -s}{\rm d} s,
  \label{eq:taylorA}
\end{align}
noting that apart from the resonant factor, the integrand is well-behaved around $s=1$ for all values of $\tilde\omega_1$ as long as $\epsilon \ll 1$. The $A_n$ are simply the coefficients of the Taylor expansion of the function $f(s)\equiv\Sigma(s)/(1-\rmi s K(1+\epsilon\tilde\omega_1-s))$ at $s=1$:
\begin{align}
    A_n = \frac{f^{(n)}(1)}{n!},
\end{align}
and for our constant size distribution (\ref{eq:sizedistsigma}) the first few terms read:
\begin{align}
    A_0=&\frac{\Sigma(1)}{1-\rmi K\epsilon\tilde\omega_1},\\
    A_1=&\frac{\rmi K(\epsilon\tilde\omega_1-1)\Sigma(1)}{(\rmi + K\epsilon\tilde\omega_1)^2},\\
    A_2=& \frac{\rmi K (K (\epsilon^2\tilde\omega_1^2 - \epsilon\tilde\omega_1 + 1) + i)}{(K \epsilon\tilde\omega_1 + \rmi)^3}.
\end{align}

If the size distribution is narrow enough so that $|1-s| \ll \epsilon|\tilde\omega_1|$ for all $s$, we can approximate the integrand by a constant:
\begin{align}
  {\rm i}\epsilon^2\int \frac{\Sigma(s)}{(1 + \epsilon\tilde\omega_1 -s)(1-{\rm i}sK\left(1 + \epsilon\tilde\omega_1 - s\right))}{\rm d} s \approx & \nonumber\\
    {\rm i}\epsilon^2(s_{\rm max}-s_{\rm min})\frac{\Sigma(1)}{\epsilon\tilde\omega_1(1-{\rm i}K\epsilon\tilde\omega_1)}= & \nonumber\\
    \frac{{\rm i}\epsilon}{\tilde\omega_1(1-{\rm i}K\epsilon\tilde\omega_1)} \approx &
    \frac{{\rm i}\epsilon}{\tilde\omega_1}. 
\end{align}
Taking $s_{\rm max}=1+\delta/2$ and $s_{\rm min}=1-\delta/2$, the condition that the integration domain is narrow enough, translates into $\delta \ll \epsilon |\tilde\omega_1|$. This is going to be our definition of a narrow size distribution.

We therefore find that, for a narrow size distribution as defined above, that $I^{(1)}= {\rm i}/\omega_1$, and therefore
\begin{align}
\tilde\omega_1 = \bar\mu^{1/2}\frac{1+{\rm i}}{2\sqrt{K}},
\label{eq:adi_monogrowth}
\end{align}
which is equivalent to the monodisperse result. This limit holds for $\delta \ll \epsilon |\tilde\omega_1|$, or
\begin{align}
  K \ll \frac{\mu}{4\delta^2}.
  \label{eq:adi_cutoffK}
\end{align}
We can recast this as
\begin{align}
  \alpha\left(\tau_{\rm max} - \tau_{\rm min}\right) \ll \frac{\omega }{k}-c_{\rm g}.
  \label{eq:ADI_quasi}
\end{align}
Therefore, the system acts as monodisperse if the perturbation in wave velocity is much larger than the width of the background dust velocity distribution. We call this regime \emph{quasi-monodisperse}. The resonance can tolerate a wider size distribution for larger dust to gas ratios and smaller wave numbers, both of which increase the perturbed wave velocity. 

\begin{figure}
  \resizebox{\hsize}{!}{\includegraphics{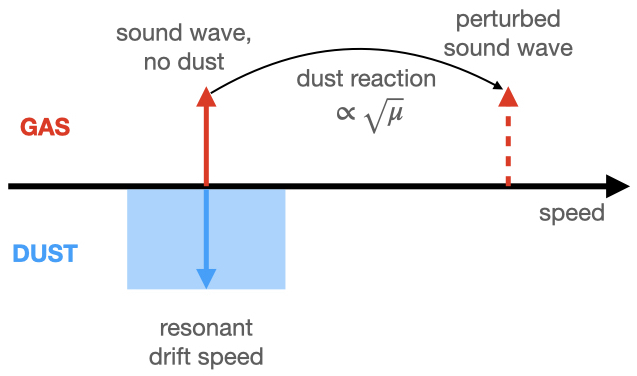}}
  \caption{Schematic representation of the quasi-monodisperse regime. The top half of the figure deals with the gas, while the bottom half deals with the dust component. The horizontal axis indicates (wave) speed. The red solid arrow denotes the propagation speed of a gas sound wave in the absence of dust, and the blue arrow indicates the resonant dust drift speed. The backreaction of the dust modifies the propagation speed of the sound wave, leading to the dashed red arrow. The blue rectangle indicates a range of drift speeds if there is a distribution of dust sizes. As long as this rectangle does not come close to the dashed red arrow, we are in the quasi-monodisperse regime.}
  \label{fig:ADI_schematic}
\end{figure}

The quasi-monodisperse regime is illustrated schematically in figure \ref{fig:ADI_schematic}. We start with a gas sound wave (the solid red arrow), propagating in the absence of dust at a speed indicated by its horizontal position. The resonant drift speed is indicated by the blue arrow, where the dust drifts at the same velocity as the propagation speed of the gas wave. The backreaction of the dust on the gas leads to a perturbed wave speed for the gas, a perturbation that is proportional to $\mu^{1/2}$ (see (\ref{eq:adi_monogrowth})). The perturbed gas wave is indicated by the dashed red arrow. A dust size distribution gives rise to a range of drift speeds, indicated by the blue rectangle. As long as this blue rectangle stays far away from the dashed red arrow, we are in the quasi-monodisperse regime, see (\ref{eq:ADI_quasi}).  

Under the conditions of a narrow size distribution, the integral in the full dispersion relation (\ref{eq:adi_disp}) can be approximated by a constant in general, resulting in the exact same dispersion relation as in the monodisperse case. This immediately means that, in the quasi-monodisperse regime, we also find the "long wavelength pressure-free mode" \citep{2018MNRAS.480.2813H} for $K \ll \mu$, with a growth rate $\propto K^{2/3}\mu^{1/3}$, and a "short wavelength resonant quasi-drift mode", with a growth rate $\propto K^{1/3}\mu^{1/3}$ \citep{2018MNRAS.480.2813H, 2024MNRAS.529..688M}. Note, however, that the latter requires $K \gg 1/\mu$ and $K \ll \mu/\delta^2$, which means that this short wavelength mode can only operate for extremely narrow size distributions that have $\delta \ll \mu$.  

\begin{figure}
  \resizebox{\hsize}{!}{\includegraphics{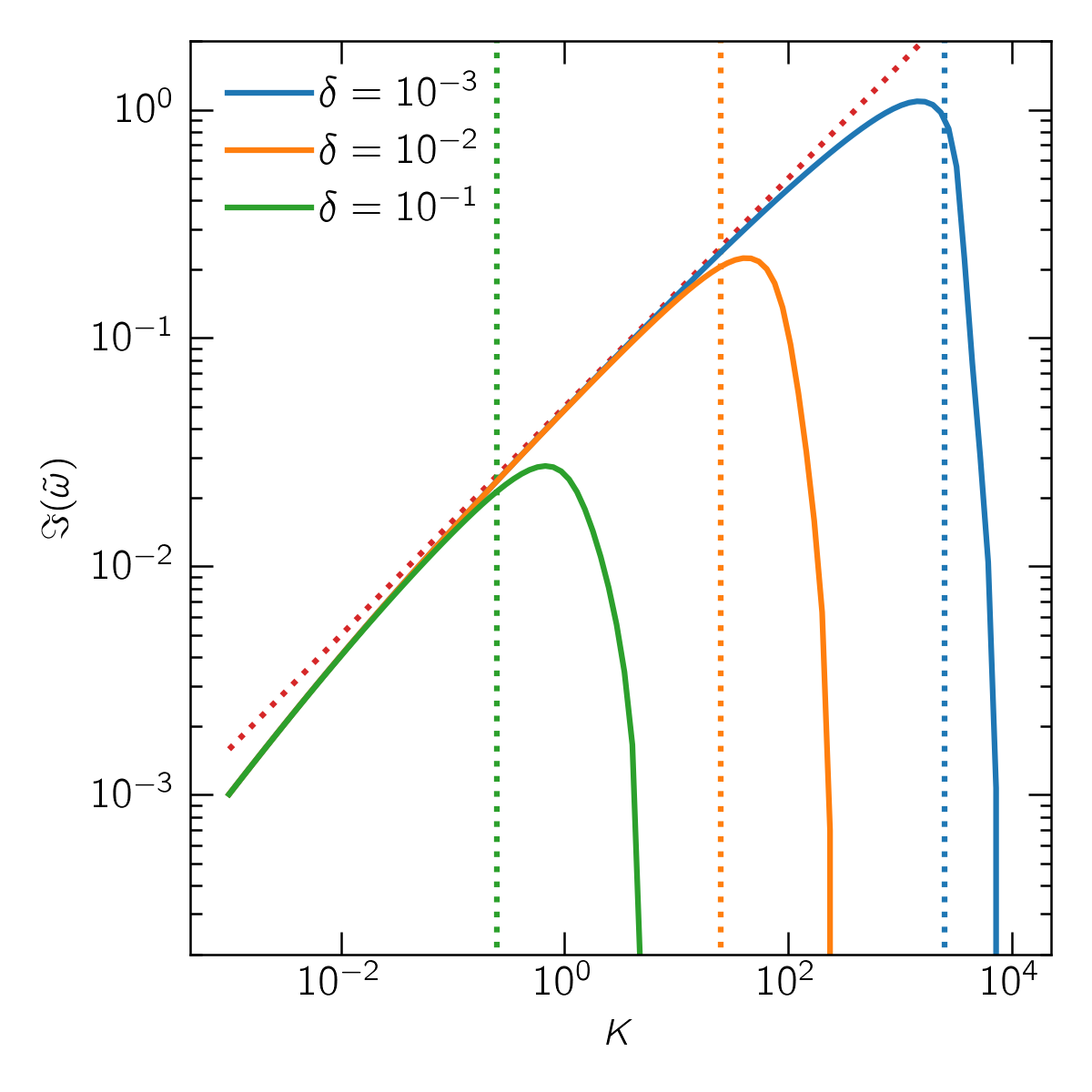}}
  \caption{Growth rate of the acoustic resonant drag instability as a function of wave number, obtained by numerically solving (\ref{eq:adi_disp}). The dust to gas ratio $\mu=0.01$, and the size density is constant, centered on the resonant size $s=1$, with width $\delta$. The red dotted line indicates the monodisperse result, while the vertical dotted lines indicate the cutoff of (\ref{eq:adi_cutoffK}).}
  \label{fig:ADI_Kmu}
\end{figure}

These results are illustrated in figure \ref{fig:ADI_Kmu}, where we consider the case with $\mu=0.01$ and the size distribution given by (\ref{eq:sizedistsigma}). The monodisperse limit, indicated by the red dotted line, shows growth at all wavenumbers, with the highest wave numbers growing fastest. In the polydisperse case, there is a cutoff at high wave numbers, which is where the system leaves the quasi-monodisperse regime. We will see below that growth is more difficult if the size distribution is too wide. For wave numbers far below the cutoff (\ref{eq:adi_cutoffK}) and $K=O(1)$, growth rates follow the monodisperse result $\propto K^{1/2}$. For $K\ll \mu$, the growth rates follow that of the long wavelength pressure-free mode $\propto K^{2/3}$. 

\subsubsection{Wide size distribution}
We now return to (\ref{eq:taylorA}) but do not make the assumption of a narrow size distribution. The series expansion around $s=1$ does not necessarily converge far away from $s=1$, depending on the value of $K$: for $K=1$, for example, we need $|s-1|<0.5$. For larger values of $K$, the size distribution needs to be more narrow in order for the series expansion to converge for all values of $s$ in the domain, while smaller values of $K$ means we can tolerate a wider size distribution. Regardless of the convergence of the series, we can always split off the resonant term involving $n=0$.  

If we do not make the assumption of a narrow size distribution, for $n=0$ in (\ref{eq:taylorA}) we get from the full integral:
\begin{align}
  \rmi\epsilon^2A_0\int \frac{1}{1 +\epsilon \tilde\omega_1-s}{\rm d} s =\nonumber\\
  -\rmi\epsilon^2A_0\int \frac{x-{\rm i}}{x^2+1}{\rm d} x =\nonumber\\
   \rmi\epsilon^2A_0\left[-\frac{1}{2}\log(1+x^2) + {\rm i} \tan^{-1}x\right]_{x_{\rm min}}^{x_{\rm max}},
   \label{eq:adi_wide_integral}
\end{align}
where
\begin{align}
    x \equiv \frac{1-s+\epsilon\Re(\tilde\omega_1)}{\Im(\tilde\omega_1)}.
\end{align}
As an aside, note that the integral diverges in the limit where the imaginary part of $\tilde\omega_1$ vanishes. This is why standard numerical integration techniques have difficulty evaluating integrals of this type, in particular for growth rates that are close to zero \citep{2021MNRAS.502.1579P}.

Define a wide size distribution by taking $s_{\rm max}=1+\Delta_R$ and $s_{\rm min}=1-\Delta_L$, with $\Delta_{L,R} \sim 1$:
\begin{align}
  \int \frac{1}{1 +\epsilon\tilde\omega_1-s}{\rm d} s =& -\frac{1}{2}\log\left(\frac{\epsilon^2\Im(\tilde\omega_1)^2+(\epsilon\Re(\tilde\omega_1)-\Delta_R)^2}{\epsilon^2\Im(\tilde\omega_1)^2 + (\epsilon\Re(\tilde\omega_1)+\Delta_L)^2}\right) \nonumber\\ 
  + {\rm i} \tan^{-1}\left(\frac{\epsilon\Re(\tilde\omega_1)-\Delta_R}{\epsilon\Im(\tilde\omega_1)}\right)- &{\rm i} \tan^{-1}\left(\frac{\epsilon\Re(\tilde\omega_1)+\Delta_L}{\epsilon\Im(\tilde\omega_1)}\right)\nonumber\\
  =& \log\left(\frac{\Delta_L}{\Delta_R}\right)-{\rm i}\pi + O(\epsilon).
\end{align}
For a wide size distribution, therefore, (\ref{eq:taylorA}) is $O(\epsilon^2)$, and we must have $\tilde\omega_1=0$, and we can subsequently neglect all $\epsilon$ terms under the integral:
\begin{align}
  \tilde\omega_2 =&\frac{\bar\mu}{2K}\int \frac{\Sigma(s)\left({\rm i}-K(1 -s)^2\right)}{(1 -s)(1-{\rm i}sK\left(1- s\right))}{\rm d} s - \frac{{\rm i}\bar \mu}{2K}.
  \label{eq:omega2}
\end{align}
The term proportional to $K$ under the integral is due to the perturbation in relative velocity, and it is straightforward to show that the contribution of this term to the imaginary part of $\omega_2$ is always negative. Together with the last term, representing the gas density perturbation, the relative velocity perturbation is therefore a stabilizing effect. The dust density perturbation has to overcome both in order for the instability to grow. 

For simplicity and avoidance of cluttered notation, we have only considered $n=0$ in (\ref{eq:taylorA}). It is straightforward to show, by defining $x=1+\epsilon\tilde\omega_1-s$:
\begin{align}
    \int \frac{(1-s)^n}{1+\epsilon\tilde\omega_1-s}\rmd s = 
    -\int \frac{(x-\epsilon\tilde\omega_1)^n}{x}\rmd x,
\end{align}
and expanding the term in the numerator, that all terms are $O(\epsilon^2)$ and that the conclusion that we need $\tilde\omega_1=0$ holds in general.

If we split off the resonant term in (\ref{eq:omega2}), we get that
\begin{align}
  \tilde\omega_2 =&
  \frac{\bar\mu}{2K}\int \frac{\rmi\Sigma(s)}{1 -s}{\rm d} s -
  \frac{\bar\mu}{2}\int \frac{\Sigma(s)}{1-{\rm i}sK\left(1- s\right)}{\rm d} s 
  - \frac{{\rm i}\bar \mu}{2K},
\end{align}
or, integrating the first term as above, and including the full frequency:
\begin{align}
\omega =& kc_{\rm g} + 
  \frac{\mu  \sigma^{(0)}(\tau_{\rm res})}{2 \rho_{\rm d}^{(0)}}\left[\rmi\log\left(\frac{\Delta_L}{\Delta_R}\right)+\pi\right]\nonumber\\
  -&\frac{\mu kc_{\rm g}}{2\rho_{\rm d}^{(0)}}\int \frac{\sigma^{(0)}({\tau_{\rm s}})}{1-{\rm i}{\tau_{\rm s}} kc_{\rm g}\left(1- {\tau_{\rm s}}/\tau_{\rm res}\right)}{\rm d} {\tau_{\rm s}}
  - \frac{{\rm i}\mu }{2 \tau_{\rm res}},
  \label{eq:adi_polygrowth}
\end{align}
with resonant stopping time $\tau_{\rm res}=c_{\rm g}/\alpha$. The contribution of the resonance (the second term on the right hand side), is proportional to the size density at the resonant stopping time, therefore increasing the importance of the resonance if the size distribution is dominated by the resonant size. In order for the resonance to promote instability, we need $\log(\Delta_L/\Delta_R)> 0$, so $\Delta_L/\Delta_R > 1$: the size distribution should not extend too far towards larger sizes. This particular effect is due to our assumption that $\Sigma$ is constant: for more general power law size distributions there can be growing modes for $\Delta_R \gg 1$. These two issues make the \adi{} sensitive to the exact size distribution: the importance of the resonant size, as well as the maximum and minimum size.  

It is important to realise that the fact that $\tilde\omega_1=0$ means that growth is now proportional to $\mu$ rather than $\sqrt{\mu}$ which is the hallmark of an RDI \citep{2018ApJ...856L..15S}. Mathematically speaking, integrating over the singularity that is the resonance gives a finite result, essentially removing the resonant nature of the instability. Physically speaking, the dust density contribution to the instability is smaller in the polydisperse case, to the extent that it now has to compete with stabilizing effects of velocity perturbations and gas density perturbations.

The remaining integral in (\ref{eq:adi_polygrowth}) contains contributions from both the (non-resonant) density and relative velocity perturbation. While the integral can be solved in terms of a complex inverse tangent function, in practice it is easier to solve the integral numerically. The non-resonant contribution to the density perturbation is usually much smaller than the resonant contribution, unless $\Delta_R \approx \Delta_L$. We will assume this is the case in what follows. 

As in the case of a narrow size distribution, there is again an important wave number dependence: the stabilizing effect of the velocity perturbations becomes stronger with increasing $K$. If we approximate the contribution of the relative velocity perturbation as
\begin{align}
 -\frac{{\rm i}K\Sigma(1)}{2}\int \frac{s\left(1 - s\right)^2}{1+s^2K^2\left(1 - s\right)^2}{\rm d} s\approx -\frac{{\rm i}K\Sigma(1)}{2}\int s\left(1 - s\right)^2{\rm d} s,
\end{align}
then the imaginary part of $\omega_2$ is, ignoring non-resonant contributions from the density perturbation, 
\begin{align}
  \Im\left(\tilde\omega_2\right)=
  \frac{\bar\mu\Sigma(1)}{2K}\log\left(\frac{\Delta_L}{\Delta_R}\right) -
              \frac{\bar\mu K\Sigma(1)}{2}\int s\left(1 - s\right)^2{\rm d} s - \frac{\bar\mu}{2K},
\end{align}
which means there is a cut-off wave number
\begin{align}
K^2<\frac{\Sigma(1)\log\left(\frac{\Delta_L}{\Delta_R}\right) - 1}{\Sigma(1)\int s\left(1 - s\right)^2{\rm d} s} = O(1),
\label{eq:cutoffKpoly}
\end{align}
which means that for $K \gg 1$ we expect no growth. This is purely because the relative velocity perturbations provide stability at small scales.  

In the limit of $K \ll \mu$ we can find a polydisperse version of the long wavelength pressure-free mode, as identified in \cite{2022MNRAS.510..110S}. Rescaling the wavenumber to $K=\bar K \epsilon^4$, with $\bar K = O(1)$, and at the same time setting $\tilde\omega = \bar \omega \epsilon^3$, we find from (\ref{eq:adi_disp}) that
\begin{align}
  \epsilon\bar\omega^2 =&
              \epsilon^3 K^2 + \bar\mu
              \int \Sigma(s)\frac{{\rm i}\bar K\bar\omega-\epsilon^{2}\bar\omega(\bar\omega-\epsilon\bar K s)^2}{(\bar\omega-\epsilon Ks)(1-{\rm i}s\epsilon^3\left(\bar\omega- \epsilon Ks\right))}{\rm d} s - {\rm i}\bar K\bar\mu.
\end{align}
At lowest order in $\epsilon$, this holds straightforwardly. At order $\epsilon$, we need
\begin{align}
  \bar\omega^3 =& {\rm i}\bar\mu \bar K^2\int \Sigma(s)s{\rm d} s 
  ={\rm i}\bar\mu \bar K^2 \frac{s_{\rm max} + s_{\rm min}}{2}.
\end{align}
A similar result was obtained by \cite{2022MNRAS.510..110S} (their appendix A). 

\begin{figure}
  \resizebox{\hsize}{!}{\includegraphics{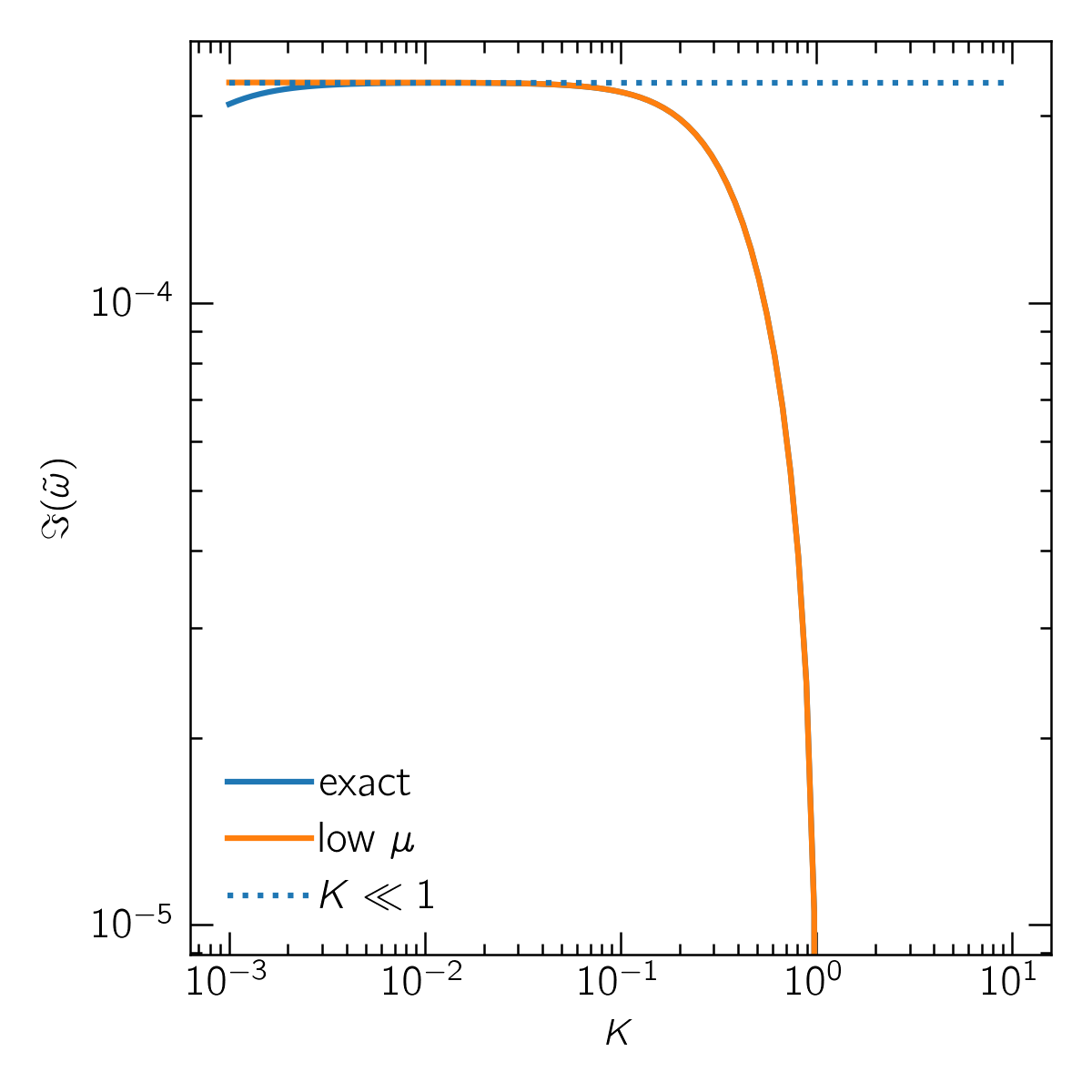}}
  \caption{Growth rate for the backward propagating sound wave, with $\omega_0=-kc_{\rm g}$, for $\mu=0.001$, $s_{\min}=0.5$ and $s_{\rm max}=1.2$. The solid blue curve represents the exact solution to (\ref{eq:adi_disp}), while the orange curve is the approximation (\ref{eq:adi_backward}) for $\mu\ll 1$, which overlaps with the blue curve everywhere except for $K \lesssim \mu$. The dotted line is the approximation (\ref{eq:adi_backward_approx}) for small $K$.}
  \label{fig:ADI_backward}
\end{figure}

\begin{figure}
  \resizebox{\hsize}{!}{\includegraphics{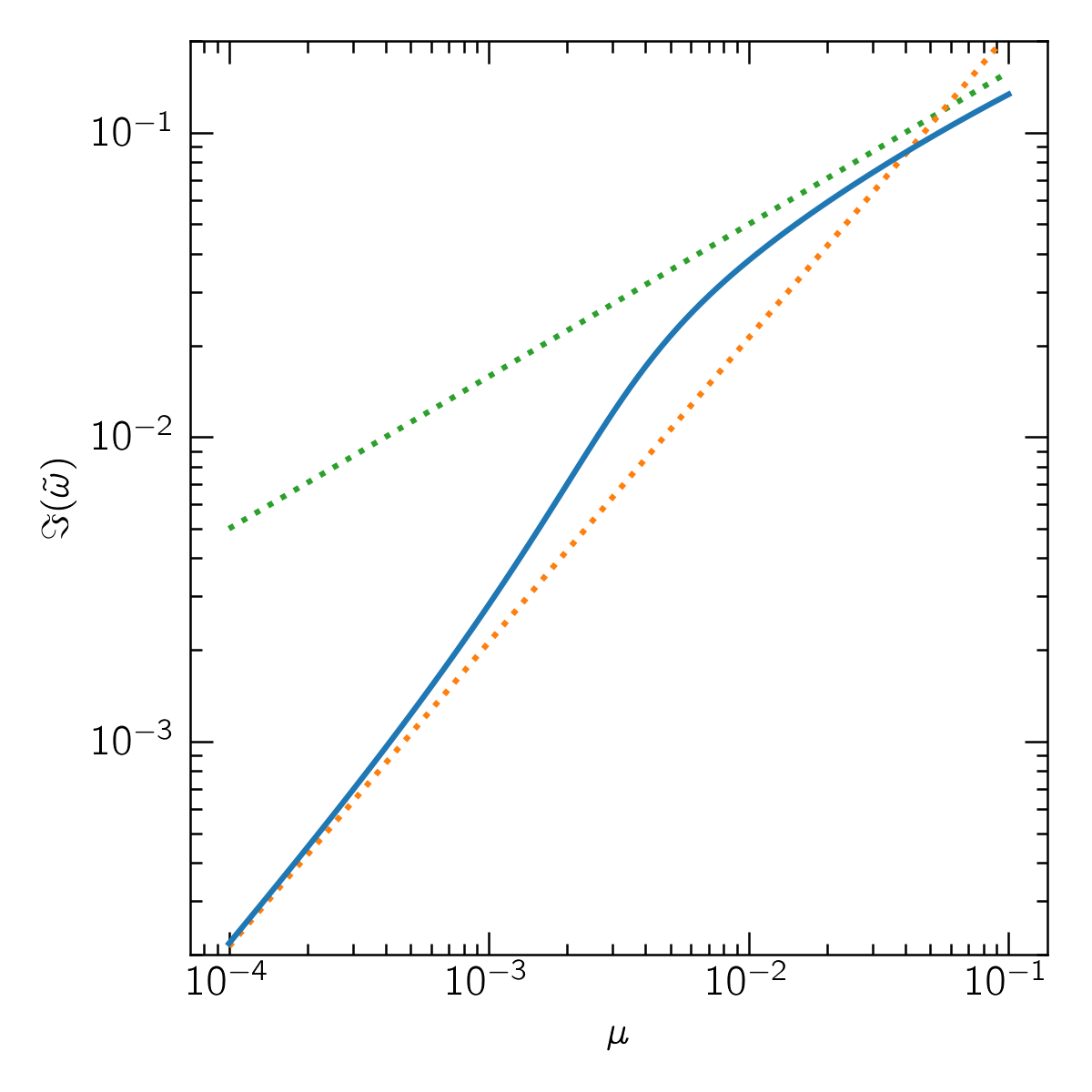}}
  \caption{Growth rate of the acoustic resonant drag instability as a function of dust to gas ratio $\mu$, calculated numerically from (\ref{eq:adi_disp}). The dimensionless wavenumber $K=1$, and the size density is constant between $s_{\rm min}=0.875$ and $s_{\rm max}=1.05$. The green dotted line indicates the monodisperse result (\ref{eq:adi_monogrowth}), while the orange dotted line indicates the polydisperse limit (\ref{eq:adi_polygrowth}).}
  \label{fig:ADI_mu}
\end{figure}

\begin{figure}
  \resizebox{\hsize}{!}{\includegraphics{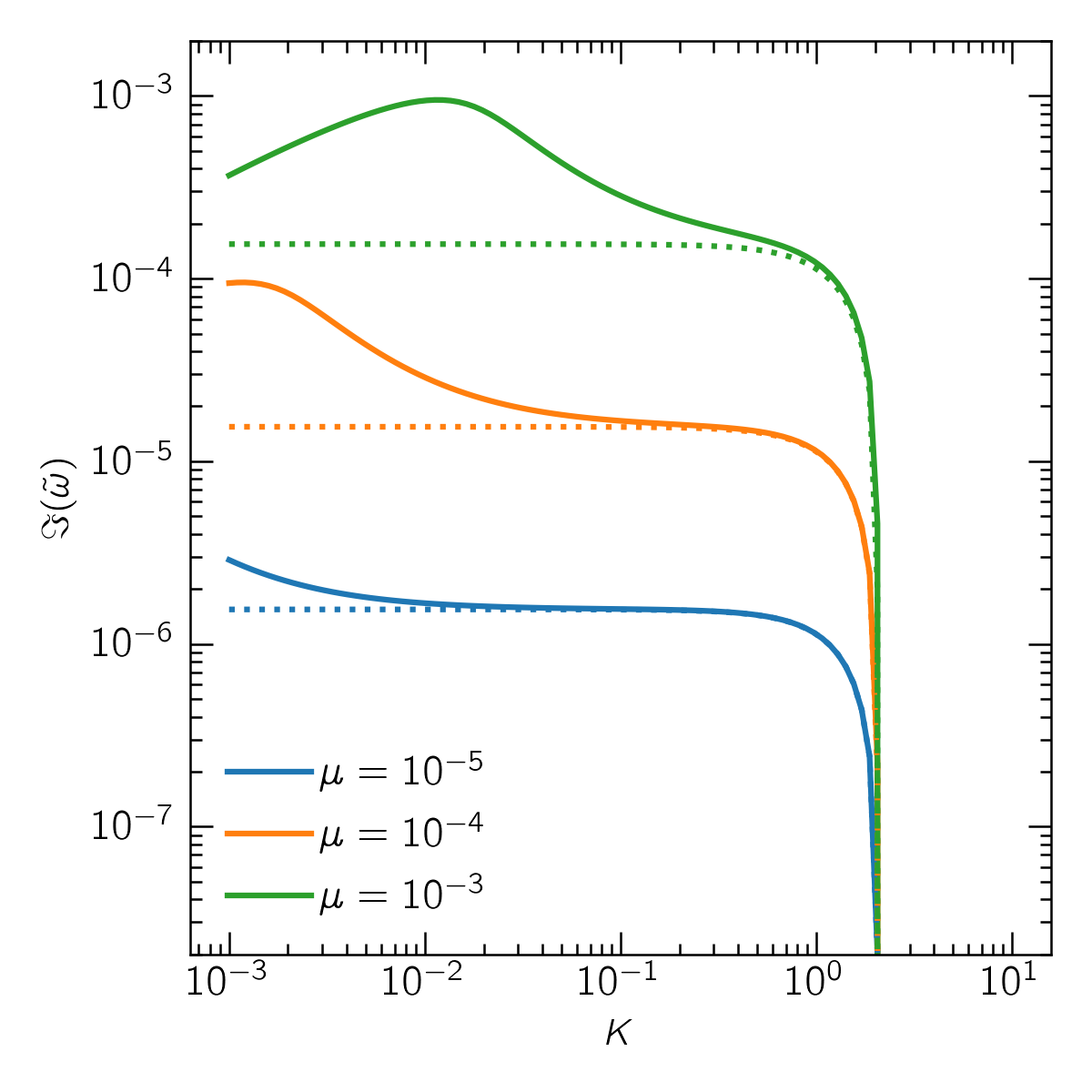}}
  \caption{Growth rate of the acoustic resonant drag instability as a function of wave number for different dust to gas ratios $\mu$, calculated numerically from (\ref{eq:adi_disp}). The size distribution is a constant between $s_{\rm min}=0.5$ and $s_{\rm max}=1.2$. The dotted curves indicate the approximation (\ref{eq:adi_polygrowth}).}
  \label{fig:ADI_asym}
\end{figure}

\subsection{Interpretation}

We summarize the results of the previous subsections using figures \ref{fig:ADI_Kmu}, \ref{fig:ADI_backward}, \ref{fig:ADI_mu} and \ref{fig:ADI_asym}. We start with the non-resonant case in figure \ref{fig:ADI_backward}. This is the case of a backward propagating soundwave with $\tilde\omega_0=-K$ going unstable. We focus on $\mu=0.001$, and a constant size distribution with $0.5 < s < 1.2$. The exact result, obtained by solving (\ref{eq:adi_disp}), indicated by the blue curve, shows growth for $K<1$ and stability for $K>1$ \citep[see also][]{2022MNRAS.510..110S}. The approximation (\ref{eq:adi_backward}) for $\mu \ll 1$, indicated by the orange curve, is indistinguishable from the exact result except for $K \lesssim \mu$, which is where our ordering scheme becomes invalid. The limit (\ref{eq:adi_backward_approx}) of $K\ll 1$ does a good job of predicting the plateau of maximum growth. The maximum growth rate is comparable to the monodisperse result for $s=1$, in which case $\tilde \omega_1=\rmi/4$ for $K \ll 1$. However, in the monodisperse case, this growth rate of $O(\mu)$ is usually dwarfed by the resonant case which has growth of $O(\sqrt{\mu})$. In the polydisperse case, however, this mode tends to dominate because of the reduced growth rates of the resonant modes \citep{2022MNRAS.510..110S}, which we explore below. 

In the previously discussed figure \ref{fig:ADI_Kmu}, we consider narrow size distributions for a dust to gas ratio $\mu=0.01$. For small $K$, the results follow the monodisperse result of (\ref{eq:adi_monogrowth}). Note that for very small wave numbers, the growth rate starts to differ from (\ref{eq:adi_monogrowth}), which represents the regime of the "long wavelength pressure-free mode" \citep{2018MNRAS.480.2813H, 2024MNRAS.529..688M}. This regime occurs for $K\ll \mu$, and occurs most readily in the quasi-monodisperse case given by (\ref{eq:adi_cutoffK}). For a wide size distribution, it is hard to enter the $K \ll \mu$ regime without entering the quasi-monodisperse regime at the same time (see also below). 

Towards larger $K$, growth disappears as we leave the quasi-monodisperse regime. In this simplified system, this means that the width of the size distribution actually sets the maximum growth rate. In reality, there is probably competition from diffusive processes killing off growth at high wave numbers. We briefly look at the effects of diffusion in section \ref{sec:visc}. As noted in \cite{2018MNRAS.480.2813H}, subsonic drift also leads to a cut-off at high wave numbers, see (\ref{eq:adi_nonres_cutoff}). For $\mu=\delta=0.01$, the cut-off occurs at a similar wave number as for monodisperse subsonic drift with $s\approx 0.9$.

In figure \ref{fig:ADI_mu} we show the growth rate at fixed $K=1$ and a size distribution that is constant between $s_{\min} = 0.875$ and $s_{\rm max}=1.05$. These limits were chosen to get the transition between quasi-monodisperse and the polydisperse regime at reasonable wave numbers. When $\mu$ is large enough, the quasi-monodisperse condition (\ref{eq:adi_cutoffK}) is satisfied, and we recover a growth rate $\propto \sqrt{\mu}$, characteristic of resonant drag instabilities. Although the size distribution is asymmetric around $s=1$, if we take $\delta=0.0875$ as an average width, from (\ref{eq:adi_cutoffK}) we expect the quasi-monodisperse regime to hold for $\mu \gg 0.03$. At lower values of $\mu$, we leave the quasi-monodisperse regime and thereby the dominance of the resonance, which leads to a growth rate $\propto \mu$. 

In figure \ref{fig:ADI_asym} we show results for a wide size distribution, which is constant between $s_{\rm min}=0.5$ and $s_{\rm max}=1.2$. Note that we need the the size distribution to be asymmetric around the resonant size $s=1$, otherwise the second order contribution from the dust density perturbation vanishes, see (\ref{eq:adi_polygrowth}). This is also where the shape of the size distribution can make a difference. At large wave numbers $K\gtrsim 1$, growth vanishes, see (\ref{eq:cutoffKpoly}). The plateau in growth towards the right side of the curves, most notably for $\mu=10^{-5}$, is the polydisperse mode, and it can be readily verified that growth in this regime is indeed $\propto \mu$. The size distributions considered in figure \ref{fig:ADI_Kmu} are too narrow for polydisperse growth to show up. The dotted curves show the relation (\ref{eq:adi_polygrowth}), which does a good job in predicting the plateau and the cut-off at high wave numbers. The bump towards the left of the curves is the tail of the quasi-monodisperse regime, which is not part of the derivation of (\ref{eq:adi_polygrowth}). If we take $\delta=0.35$ for this size distribution, from (\ref{eq:adi_cutoffK}) we expect the quasi-monodisperse regime to hold for $K \ll 2\mu$, consistent with what is seen in figure \ref{fig:ADI_asym}. 
 
In summary, the polydisperse version of the \adi{} is different from its monodisperse counterpart in two important ways. First, a distribution of dust sizes lead to a cutoff of growth at high wave numbers, even for very narrow size distributions. Second, for wider size distributions, there exists an intermediate range of wave numbers, where the resonance plays only a minor role, as the integration over size dilutes its contribution. Here, the growth rates scale with $\mu$, rather than $\sqrt{\mu}$, which is the case for monodisperse RDIs. In other words, a wide size distribution converts resonant growth into ordinary growth, with usually reduced growth rates for $\mu \ll 1$. At smaller wave numbers, we find the quasi-monodisperse regime, which behaves similar to the monodisperse \adi{}. 

\subsection{Acoustic Drag Instability with diffusion}
\label{sec:visc}

Before we conclude, we briefly look at diffusive effects by including gas viscosity and dust diffusion. We only consider the monodisperse version of the \adi{}. It is expected that diffusion will stabilise the instability at large wave numbers, and it is interesting to compare the diffusive cut-off to the cut-off due to having a size distribution. Realistic levels of viscosity and associated dust diffusion will strongly depend on the background state, in particular whether the underlying gas flow is turbulent \citep[see e.g.][for the case of AGB star winds]{2006A&A...452..537W}. Here, we do not aim at modeling a particular physical situation in terms of the input level of viscosity, but rather ask the question at what level of viscosity do dissipative effects associated with gas viscosity and dust diffusion compete with effects due to the finite with of the size distribution discussed in the previous subsections. 

The monodisperse equations, including gas viscosity and dust diffusion, read, adopting the formulation of \cite{2021ApJ...907...64L}:
\begin{align}
  \partial_t\rho_{\rm g} + \partial_x(\rho_{\rm g} v_{\rm g})=&0,\\
  \partial_t v_{\rm g} + v_{\rm g} \partial_xv_{\rm g} =& -\mu \alpha -
  \frac{c_{\rm g}^2\partial_x\rho_{\rm g}}{\rho_{\rm g}} +
  \frac{\rho_{\rm d}}{\rho_{\rm g}}\frac{u-v_{\rm g}}{\tau_{\rm s}}\nonumber\\
  +& \frac{4}{3\rho_{\rm g}}\nu\partial_x\left(\rho_{\rm g}\partial_xv_{\rm g}\right),\\
   \partial_t\rho_{\rm d} + \partial_x(\rho_{\rm d}
  u)=&D\partial_x\left((\rho_{\rm g} + \rho_{\rm
       d})\partial_x\left(\frac{\rho_{\rm d}}{\rho_{\rm d}+\rho_{\rm g}}\right)\right),\\
  \partial_t u+ u\partial_xu =& \alpha-\frac{u-v_{\rm g}}{\tau_{\rm s}},
\end{align}
where $\nu$ is the gas kinematic viscosity and $D$ the dust diffusion coefficient. The background flow is constant in space and time, and exactly the same as the inviscid case. Linear perturbations are governed by:
\begin{align}
  \partial_t\mathcal{G}+ \partial_xv_{\rm g}^{(1)}=&0,\\
  \partial_t v_{\rm g}^{(1)}
  =& -c_{\rm g}^2\partial_x\mathcal{G} + \mu\frac{u^{(1)}-v_{\rm
     g}^{(1)}}{\tau_{\rm s}} \nonumber\\
     +& \mu\left(\mathcal{D}- \mathcal{G}\right) \frac{u^{(0)}}{\tau_{\rm s}} + \frac{4\nu}{3}\partial_x^2v_{\rm g}^{(1)},\\
   \partial_t\mathcal{D} + \partial_xu^{(1)}
  + u^{(0)}\partial_x\mathcal{D}=&\frac{D}{\mu+ 1}\partial_x^2\left(\mathcal{D}- \mathcal{G}\right),\\
  \partial_t u^{(1)}+ u^{(0)}\partial_xu^{(1)} =& -\frac{u^{(1)}-v_{\rm g}^{(1)}}{\tau_{\rm s}},
\end{align}
with $\mathcal{G}=\rho_{\rm g}^{(1)}/\rho_{\rm g}^{(0)}$ and
$\mathcal{D}=\rho_{\rm d}^{(1)}/\rho_{\rm d}^{(0)}$. Taking perturbations
$\propto \exp(\rmi k x - \rmi \omega t)$ as before:
\begin{align}
  k \hat v=&\omega \hat{\mathcal{G}},\\
   c_{\rm g}^2 k \hat{\mathcal{G}} + \rmi \mu\frac{\hat u-\hat
     v}{\tau_{\rm s}} + \rmi \mu \alpha\left(\hat{\mathcal{D}}-
     \hat{\mathcal{G}}\right) - \frac{4\rmi\nu}{3}k^2\hat v =& \omega \hat v,\\
     k \hat u
  + \alpha \tau_{\rm s} k \hat{\mathcal{D}} -\frac{\rmi D }{ \mu+
  1}k^2\left(\hat{\mathcal{D}}- \hat{\mathcal{G}}\right) =& \omega \hat{\mathcal{D}} ,\\
   \alpha \tau_{\rm s}k\hat u-\rmi \frac{\hat
                                                      u-\hat v}{\tau_{\rm s}}=&\omega \hat u,
\end{align}
This is a straightforward eigenvalue problem with matrix:
\begin{align}
  \mathsf{A} = \left(\begin{array}{cccc}
                       0 & k & 0 & 0\\
                       c_{\rm g}^2k - \rmi \mu \alpha & -\frac{\rmi \mu}{\tau_{\rm
                                                   s}} - \frac{4\rmi \nu
                                                   k^2}{3} & \rmi \mu \alpha
                                                           & \rmi
                                                             \mu/\tau_{\rm
                                                             s}\\
                       \frac{\rmi Dk^2}{1+\mu} & 0 & \alpha\tau_{\rm s}k-\frac{\rmi
                                              Dk^2}{1+\mu} & k\\
                       0 & \rmi/\tau_{\rm s} & 0 & \alpha\tau_{\rm s} k -
                                                   \frac{\rmi}{\tau_{\rm s}}
                     \end{array}\right)
\label{eq:eigenvalue_visc}
\end{align}
We can construct a non-dimensional version, using a length scale $c_{\rm g}^2/\alpha$ and time scale $c_{\rm g}/\alpha$:
\begin{align}
  K\hat V =&  \Omega \hat{\mathcal{G}},\\
   K\hat{\mathcal{G}} + \rmi \mu\frac{\hat U-\hat
     V}{s} + \rmi \mu \left(\hat{\mathcal{D}}-
     \hat{\mathcal{G}}\right) - \frac{4\rmi\alpha_{\rm g}}{3}K^2\hat V =& \Omega \hat V,\\
     K \hat U
  + s K\hat{\mathcal{D}} -\frac{\rmi \alpha_{\rm d}}{\mu+
  1}K^2\left(\hat{\mathcal{D}}- \hat{\mathcal{G}}\right) =& \Omega \hat{\mathcal{D}} ,\\
    s K\hat U-\rmi \frac{\hat
                                                      U-\hat V}{s}=&\Omega \hat U,
\end{align}
with nondimensional diffusion coefficients\footnote{Here we use the convention in the accretion disc community of designating the non-dimensional viscosity parameter with the Greek letter $\alpha$, in our case not to be confused with the acceleration applied to gas and dust to generate drift.} $\alpha_{\rm g} = \nu \alpha/c_{\rm g}^3$ and $\alpha_{\rm d}=D \alpha/c_{\rm g}^3$. 

\begin{figure}
  \resizebox{\hsize}{!}{\includegraphics{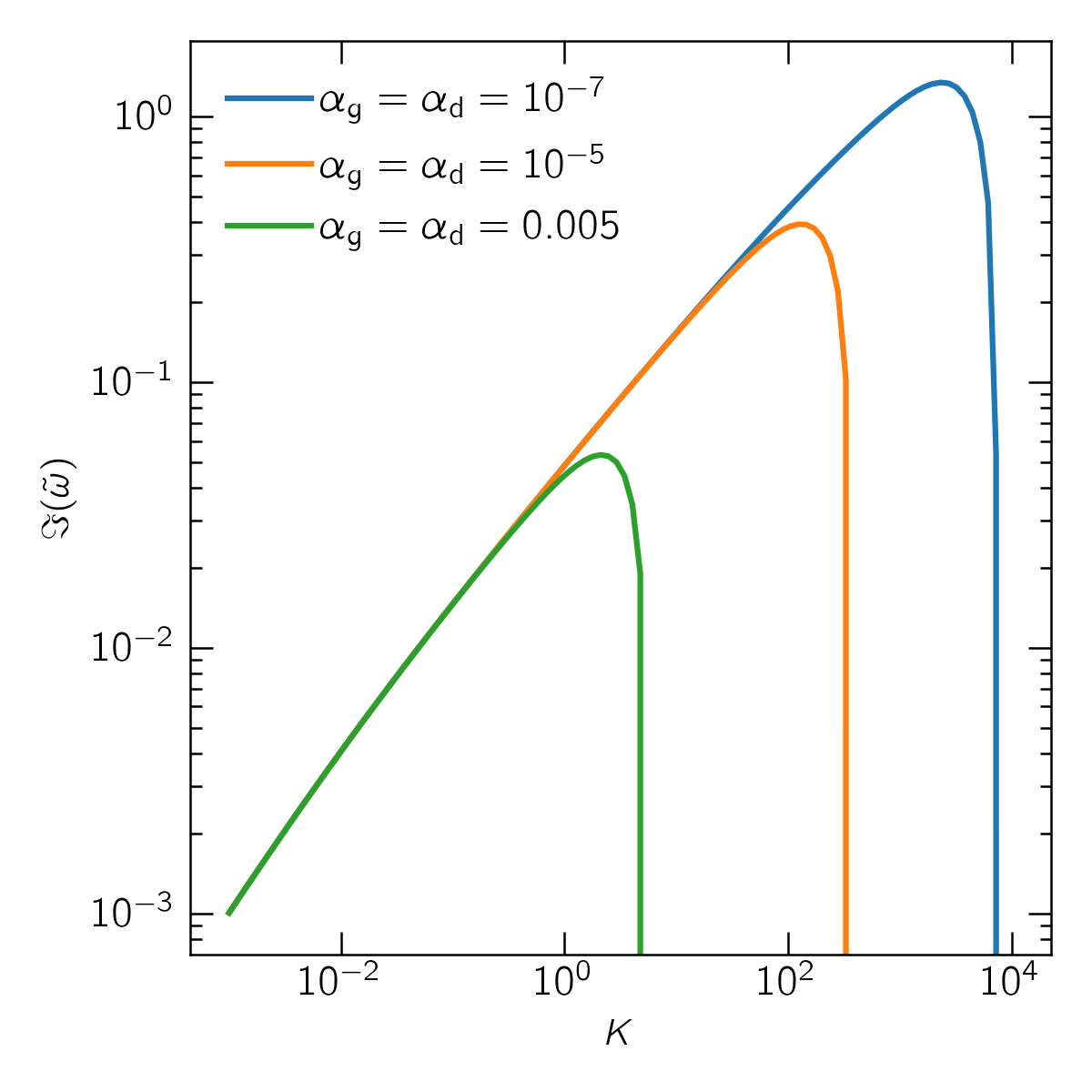}}
  \caption{Growth rate of the monodisperse acoustic resonant drag instability as a function of wave number, including gas viscosity and dust diffusion, calculated from the eigenvalues of (\ref{eq:eigenvalue_visc}). The dust to gas ratio $\mu=0.01$.}
  \label{fig:ADI_visc}
\end{figure}

The resulting growth rates for $\mu=0.01$ are shown in figure \ref{fig:ADI_visc}, for three different levels of diffusion. In all cases, we focus on the resonant case with $s=1$. For simplicity, we keep $\alpha_{\rm g}=\alpha_{\rm d}$, noting that dust diffusion is far more important than gas viscosity for damping the unstable modes. This choice corresponds to a Schmidt number of unity. The levels of viscosity were chosen so that the cutoff wave numbers roughly match those of figure \ref{fig:ADI_Kmu}. We see that a size distribution of width $\delta=0.1$ leads to a similar cutoff wavenumber as a dimensionless diffusion coefficient of $\alpha_{\rm g}=\alpha_{\rm d}=0.005$. In other words, for levels of diffusion lower than $\alpha_{\rm g}=\alpha_{\rm d}=0.005$, it is the width of the size distribution that determines the maximum growth rate, rather than diffusion. Narrower size distributions need lower levels of diffusion to be dominant. 

\section{Discussion and conclusion}
\label{sec:conclusion}

We have presented linear calculations of the acoustic resonant drag instability in the polydisperse regime. We have shown that a quasi-monodisperse regime exists, with growth rates comparable to the resonant monodisperse case, if the relative width of the size distribution around the resonant size is small:
\begin{align}
    \delta = \frac{\tau_{\rm s, max}-\tau_{\rm s, min}}{\tau_{\rm res}} \ll \frac{1}{2}\sqrt{\frac{\mu}{kc_{\rm g}\tau_{\rm res}}}.
\end{align}
There is a straightforward correspondence between the analysis presented here, and the discussion on detuning of the resonance in \cite{2024MNRAS.529..688M}, their appendix A. They studied how the growth rate varies just away from the resonance, possibly because the resonant modes do not fit exactly in the numerical domain. Our relative width of the size distribution, $\delta$, plays the role of the detuning parameter. A narrow size distribution corresponds to small detuning, while a wide size distribution corresponds to large detuning. The effect of this 'detuning' is strongest for the highest wave numbers, which is also where the monodisperse \adi{} grows fastest. The fastest growing monodisperse modes require extremely narrow size distributions to survive in the polydisperse case. 

For wider size distributions, genuinely polydisperse modes exist. Realistic widths of the dust size distribution can for example be based on the ISM: \cite{2022MNRAS.510..110S} take a minimum size of $5$ ${\rm nm}$ and a maximum size of $0.25$ $\mu m$, based on the work of \cite{1977ApJ...217..425M}. This gives a spread in sizes, and, for our simple drag law, stopping times of two orders of magnitude, or $\delta \approx 1$, definitely outside the quasi-monodisperse regime. Moreover, the true ISM size distribution may well include larger sizes \citep[e.g.][]{2001ApJ...548..296W}. In dynamical environments like AGB star winds it is more difficult to establish a size distribution, but hydrodynamical simulations including dust nucleation find that dust particles can grow rapidly over at least an order of magnitude \citep{2001A&A...371..205S}, suggestive of a wide size distribution in these environments as well. 

Even though growing polydisperse modes exist, their growth rates are $\propto \mu$ rather than $\propto \sqrt{\mu}$ for the corresponding monodisperse case, which means they usually grow slower for $\mu \ll 1$. Integrating over the resonance removes the resonant nature of the instability, but given the right size distribution, the contribution of the resonance can still promote growth. For the simple, constant size distribution considered here, we found that the size distribution should be asymmetric on either side of the resonant size. 

\begin{figure}
  \resizebox{\hsize}{!}{\includegraphics{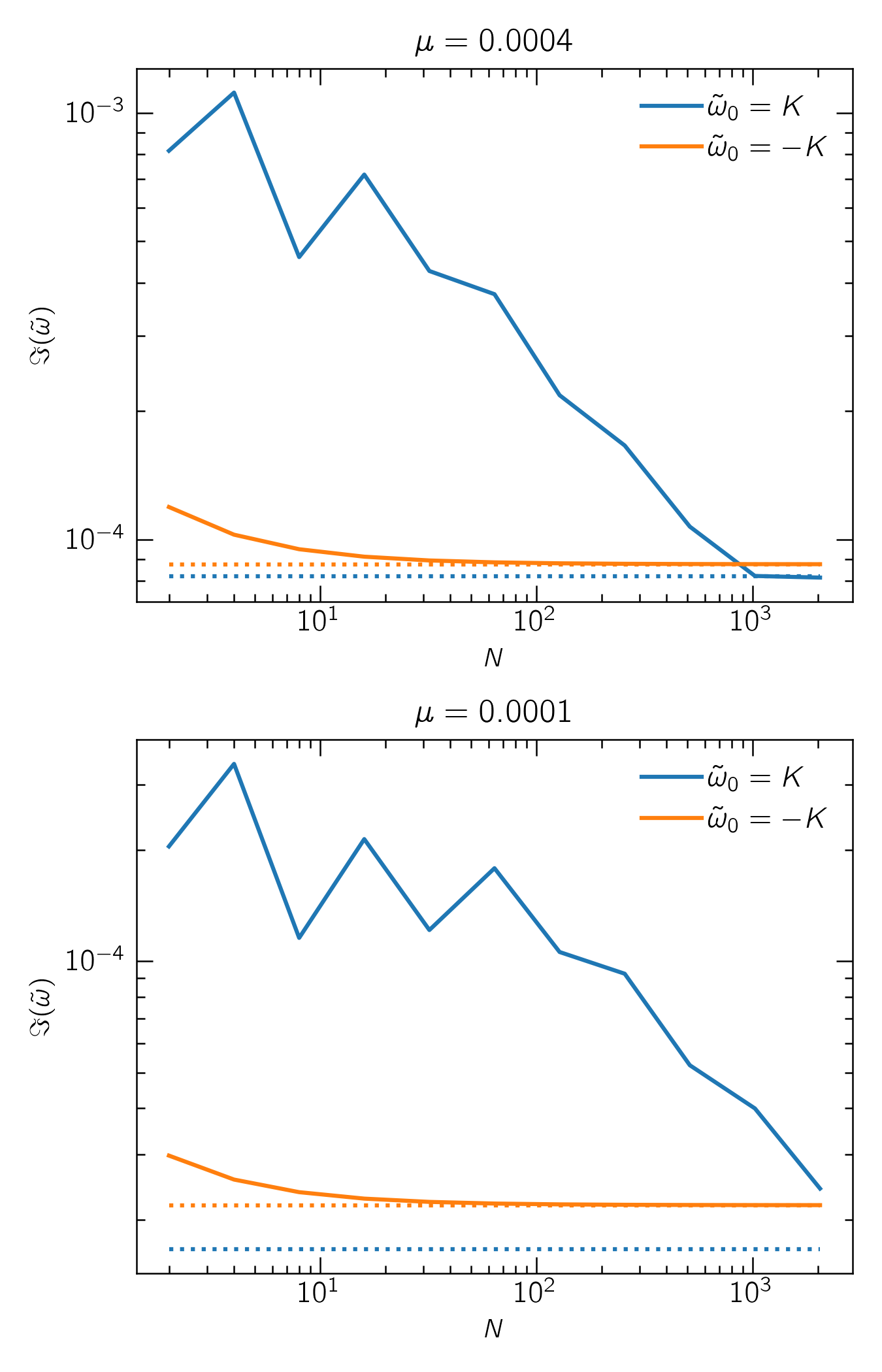}}
  \caption{Convergence of the direct method with number of dust species, for a case with $s_{\rm min}=0.5$ and $s_{\rm max}=1.2$, all at $K=0.1$. Top panel: $\mu=0.0004$, bottom panel: $\mu=0.0001$. The blue curves indicate a mode with unperturbed frequency $\tilde \omega_0=K$, which contains the resonance, and the orange curves a non-resonant mode with $\tilde\omega_0=-K$. The dotted lines indicate the result of (\ref{eq:adi_disp}).}
  \label{fig:ADI_conv}
\end{figure}

We have focused on the case of a continuous size distribution. That is, we have never explicitly discretized the integral appearing in the backreaction term in the gas momentum equation. As a result, the dispersion relation (\ref{eq:adi_disp}) contains an integral, which makes numerically solving for the frequencies not easy. A different approach is to discretize the integral from the start, and solve the resulting matrix eigenvalue problem with standard techniques \citep{2021MNRAS.502.1469M, 2022MNRAS.510..110S}. In the limit where the number of dust species $N$ goes to infinity, one expects the frequencies found to converge to the results obtained from (\ref{eq:adi_disp}). However, as noted in \cite{2021MNRAS.502.1469M} for the streaming instability, if the integral is nearly singular, this approach of discretizing the integral leads to extremely slow convergence. In figure \ref{fig:ADI_conv}, we illustrate the problem for the acoustic drag instability. The orange curves indicate the growth rates for the non-resonant backward propagating sound wave, and here we find that for $N>100$ the results are indistinguishable from the continuum case. In the case where the integration domain includes the resonance, which happens for the blue curves that have $\tilde\omega_0=K$, convergence is much harder to obtain. The top panel shows that we reach convergence for $\mu=0.0004$ at $N > 1024$, but if we lower the dust to gas ratio to $0.0001$ the result does not converge for $N$ up to $2048$. Over the whole range of $N$ considered, the fastest growing mode is not the backward propagating sound wave, but rather a spuriously growing resonant wave. Note that, for the relatively narrow size distribution chosen here, we need the dust to gas ratio to remain small or we enter the quasi-monodisperse regime (see figure \ref{fig:ADI_asym}). These spuriously growing modes could pose a problem for numerical simulations, for which it is usually too expensive to include 1000s of dust species. Even more, when we increase $K$ so that there should be no growing modes in the polydisperse regime (see figures \ref{fig:ADI_backward} and \ref{fig:ADI_asym}), the direct matrix method finds spurious growing modes with growth rates $\Im(\tilde\omega) \sim \mu$ up to $N=2048$. These can easily come to dominate a numerical simulation.  

Several important simplifications were made to make the problem tractable. First of all, we have only considered very simple (i.e. constant) size distributions. While most results are likely to be insensitive to the exact form of the size distribution (i.e. the existence of the quasi-monodisperse regime and polydisperse modes), in some cases details will matter. In particular, the contribution of the resonance to the total growth rate depends on the limits of the size distribution, see (\ref{eq:adi_polygrowth}). More complicated size distributions, either inherited from the ISM or self-consistently generated through nucleation and collisional growth, are likely to affect the growth rates of the acoustic resonant drag instability.

We have considered only a very simple drag law with constant stopping time. A more general form of Epstein drag was given in \cite{2018MNRAS.480.2813H}, where the stopping time depends both on gas density and relative velocity between gas and dust. This form was also used in \cite{2022MNRAS.510..110S}. For the fast drift necessary to trigger the \adi{}, in particular the dependence of the stopping time on relative velocity will be important. A comparison in the monodisperse case was made in \cite{2024MNRAS.529..688M}, but for the polydisperse case there is still work to be done.    

Despite the differences in setup, our results are in agreement with appendix A3 of \cite{2022MNRAS.510..110S}. In particular the existence of the unstable backward propagating sound wave for small $K$, and its dominance over the resonant mode (see figure \ref{fig:ADI_conv}). With our continuum approach, we were able to study the resonant modes in more detail. Apart from the constant stopping time mentioned above, another difference with the work of \cite{2022MNRAS.510..110S} is that we have worked in one spatial dimension only, greatly simplifying the analysis. As was mentioned in \cite{2024MNRAS.529..688M}, the physics of the instability is well-captured by a one-dimensional approach. 

In conclusion, for realistically wide size distributions, the acoustic resonant drag instability will mostly show non-resonant growth in the limit $\mu \ll 1$. The highest wave number modes are rendered stable by having a size distribution, which takes the maximum growth rate down substantially, much like diffusion. Therefore, the acoustic resonant drag instability survives in the polydisperse regime, but with growth rates that are typically $\Im(\omega) \sim \mu \alpha/c_{\rm g}$, which is a fraction $\sim \sqrt{\mu /K}$ of the monodisperse growth rate. \\

\noindent {\tiny \textit{Data Availability.} All data used to create figures in this work are available at \url{https://doi.org/10.4121/df27e782-0a37-43a9-8fdd-bd4081b96dfb}}

\begin{acknowledgements}
We thank the anonymous referee for a thorough and insightful report. This project has received funding from the European Research Council (ERC) under the
European Union’s Horizon Europe research and innovation programme (Grant Agreement No. 101054502). This work made use of several open-source software packages. We acknowledge \texttt{numpy} \citep{Harris2020Natur.585..357H}, \texttt{matplotlib} \citep{Hunter2007CSE.....9...90H}, \texttt{scipy} \citep{Virtanen2020NatMe..17..261V}, and \texttt{psitools} \citep{2021MNRAS.502.1469M}. We acknowledge 4TU.ResearchData for supporting open access to research data.
\end{acknowledgements}

\bibliographystyle{aa} 
\bibliography{rdi} 

\end{document}